\documentstyle[preprint,aps,floats]{revtex}
\tightenlines
\input psfig.tex
\begin{document}
\def\ba{\begin{eqnarray}}
\def\ea{\end{eqnarray}}
\def\be{\begin{equation}}
\def\ee{\end{equation}}
\def\({\left(}
\def\){\right)}
\def\[{\left[}
\def\]{\right]}
\def\lagrange {{\cal L}}
\def\del {\nabla}
\def\d {\partial}
\def\Tr{{\rm Tr}}
\def\half{{1\over 2}}
\def\fourth{{1\over 8}}
\def\bibi{\bibitem}
\def\S{{\cal S}}
\def\xx{\mbox{\boldmath $x$}}
\newcommand{\labeq}[1] {\label{eq:#1}}
\newcommand{\eqn}[1] {(\ref{eq:#1})}
\newcommand{\labfig}[1] {\label{fig:#1}}
\newcommand{\fig}[1] {\ref{fig:#1}}
\def\gsim{ \lower .75ex \hbox{$\sim$} \llap{\raise .27ex \hbox{$>$}} }
\def\lsim{ \lower .75ex \hbox{$\sim$} \llap{\raise .27ex \hbox{$<$}} }
\newcommand\bigdot[1] {\stackrel{\mbox{{\huge .}}}{#1}}
\newcommand\bigddot[1] {\stackrel{\mbox{{\huge ..}}}{#1}}
\title{Comment on `Quantum Creation of an Open Universe', by Andrei Linde
} 
\author{S.W. Hawking\thanks{email:S.W.Hawking@damtp.cam.ac.uk} 
and Neil
Turok\thanks{email:N.G.Turok@damtp.cam.ac.uk}}
\address{
DAMTP, Silver St, Cambridge, CB3 9EW, U.K.}
\date\today 
\maketitle

\begin{abstract}
We comment on Linde's claim that one should change the sign in the
action for a Euclidean instanton in quantum cosmology, 
resulting in the formula $P \sim e^{+S}$ for the probability of
various classical universes. 
There are serious problems with doing so. If one reverses the sign of
the action of both the instanton and the fluctuations, the latter are
unsupressed and the calculation becomes meaningless. So for
a sensible result one would have to reverse the sign of the
action for the background, while leaving the sign for the perturbations
fixed by the usual Wick rotation. The problem with this
approach is that there is no invariant way to split a given
four geometry into background plus perturbations. So the
prescription would have to violate general coordinate invariance. 
There are other indications that a sign change is problematic. 
With the choice $P \sim e^{+S}$ the 
nucleation of primordial black holes during inflation 
is unsuppressed, with a disastrous resulting cosmology.
We regard these as compelling arguments for adhering to  the
usual sign given by the Wick rotation.
\end{abstract}
\vskip .3in

In a recent letter, we pointed out the existence of new finite 
action instanton solutions describing the birth of
open inflationary universes 
according to the Hartle-Hawking 
no boundary proposal. Linde has 
written a response in which he claims that the Hartle-Hawking 
calculation of the probability for classical universes is 
wrong, and that the expresssion 
\be
 P \sim e^{-S_E(i)}
\labeq{esign}
\ee
for the probability $P$ in terms of the Euclidean action for 
the instanton solution $S_E(i)$ should be replaced by 
\be
 P \sim e^{+S_E(i)}.
\labeq{lsign}
\ee
Since the Euclidean action is very large and negative ($S_E(i) \sim - 10^8$ typically)
for solutions of
the type we describe, the difference between these two formulae is 
extremely significant.  What hope for theory if we cannot resolve 
disagreements of this order!

Let us explain where these formulae come from. One starts from the 
full Lorentzian path integral for quantum gravity coupled to a scalar field,
\be
\int [d g] [d\phi ] e^{i S[g,\phi]}
\labeq{pint}
\ee
which in principle defines all correlation functions 
of physical observables. Unfortunately the integrand is 
highly oscillatory for large field values, and an additional
prescription is needed to evaluate it.
The prescription suggested by Hartle and Hawking was to perform the 
the analytic continuation to Euclidean time, $t_E=it$, 
and to continue the metric to a compact Euclidean metric. The sign of the 
Wick rotation that is involved is fixed by the requirement that
non-gravitational physics be correctly reproduced, because the other sign
would produce an action for non-gravitational 
field fluctuations that was unbounded below. So (\ref{eq:pint}) becomes
\be
\int [d g] [d\phi ] e^{- S_E[g,\phi]}.
\labeq{peint}
\ee

Having performed the Wick rotation, 
we now try to evaluate it. The only way we know how to do this is 
to use the saddle point method. That is we find a stationary point
of the action, i.e. a solution to the classical Euclidean equations, 
and expand around it. We obtain
\be
S_E \approx S_0+S_2+...
\labeq{expansion}
\ee
where $S_0=S_E(i)$ is the action of the classical solution (the instanton) and
$S_2$ is the action for the fluctuations. One computes the 
fluctuations by performing the Gaussian integral with the 
measure exp($-{S_2}$).
It is very important that $S_2$ is 
positive so that the 
the fluctuations 
about the background classical solution are suppressed. As is well known, the 
Euclidean action for gravity alone is not positive definite, so the 
positivity of 
$S_2$ is not guaranteed, and has to be checked 
for the particular classical background  in question. In the inflationary 
example $S_2$ is known to be positive \cite{mukh}. 
Physically this corresponds to the fact that the classical background 
is not gravitationally unstable.

Let us turn to Linde's paper. He would like to reverse the sign in 
the exponent, turning (\ref{eq:lsign}) into (\ref{eq:esign}). 
This
is because the Euclidean action for the instanton 
$S_E(i) \sim  - M_{Pl}^4 /V(\phi_0)$ 
where $M_{Pl}$ is the Planck mass and $\phi_0$ the initial value of
the scalar field.  Values of the scalar field giving small values for
the potential $V(\phi_0)$ give a large negative action, and are thus
favoured. Obviously, changing the sign of the action will instead mean 
that these are strongly disfavoured, and make large initial values of
the scalar field more likely. Whilst this improves the prospects 
for obtaining large amounts of inflation,  
we do not believe the sign change is tenable. If one treats 
background and perturbations together, 
a change in the sign of $S_0
= S_E(i)$ is accompanied by a change in the sign of 
$S_2$. But this is disastrous - the fluctuations are left unsuppressed
and the description of the spacetime as a classical background
with small fluctuations breaks down.

One could try to treat background and fluctuations separately, 
by performing Wick rotations of the opposite sign on them.
However the problem is that there is no coordinate invariant way
to separate the two. So any such prescription would have to 
violate general coordinate invariance. 

Problems occur with changing the sign of the action in nonperturbative
contexts too. For example, 
calculations by Bousso and one of us \cite{BH} have shown that
if one adopted Linde's prescription 
the creation of universes with large numbers of black holes would
have been favoured and their mass would have dominated the
energy density,
leaving the universe without a radiation dominated era.

Linde gives another, intuitive, argument against using the standard 
sign for the Euclidean action.
He argues that the entropy ${\cal S} $
of de Sitter space is given by a quarter of its horizon area. 
This quantity is accurately approximated by $ {\cal S} \approx -S_E(i)$, 
the negative of the action for the Euclidean instanton. He then argues that
``it seems natural to expect that the emergence of a complicated object of
large entropy must be suppressed by exp$(-{\cal S})$''. We find this hard to
understand.
The formula probability $\propto $ exp$(+{\cal S})$
is the foundation of statistical physics. Likewise if one pictures
the formation of the universe as the endpoint of some process, the rate is
proportional to the phase 
space available in the final state, again given by exp$(+{\cal S})$.
His intuitive argument seems to us to support rather than
contradict the sign we have adopted.

In summary, changing the sign of the Euclidean 
action 
is not something one can do without negative 
repercussions. If the sign happens to 
disfavour large amounts of inflation, we prefer to
face up to that problem, as in \cite{HT}. Possible  solutions 
include a) accepting that we live in a universe
on the tail of the distribution, possibly for anthropic reasons 
or b) exploring open inflationary continuations of the type we 
proposed in the context of more fundamental theories of
quantum gravity, such as supergravity or M-theory, to see whether
large amounts of inflation are favoured for other reasons (one candidate
such mechanism was mentioned in  \cite{HT}).

\end{document}